# 2D and 3D Polar Plume Analysis from the Three Vantage Positions of STEREO/EUVI A, B, and SOHO/EIT

**Judith de Patoul**[1,2] · **Bernd Inhester**[1] · **Li Feng**[1,3] · **Thomas Wiegelmann**[1]

**Abstract** Polar plumes are seen as elongated objects starting at the solar polar regions. Here, we analyze these objects from a sequence of images taken simultaneously by the three spacecraft telescopes STEREO/EUVI A and B, and SOHO/EIT. We establish a method capable of automatically identifying plumes in solar EUV images close to the limb at $1.01-1.39\,R_\odot$ in order to study their temporal evolution. This plume-identification method is based on a multiscale Hough-wavelet analysis. Then two methods to determined their 3D localization and structure are discussed: First, tomography using the filtered back-projection and including the differential rotation of the Sun and, secondly, conventional stereoscopic triangulation. We show that tomography and stereoscopy are complementary to study polar plumes. We also show that this systematic 2D identification and the proposed methods of 3D reconstruction are well suited, on one hand, to identify plumes individually and on the other hand, to analyze the distribution of plumes and inter-plume regions. Finally, the results are discussed focusing on the plume position with their cross-section area.

**Keywords:** Corona: Quiet — Corona: Structures — Coronal Holes

## 1. Introduction

Solar polar plumes are faint, elongated structures above polar coronal holes. They consist of denser and cooler plasma compared to the surrounding. Studying plumes and inter-plume regions is of great interest for the understanding of the acceleration of the fast component of the solar wind (Teriaca *et al.*, 2003). For a long time plumes have been observed and studied in white-light during solar eclipses. Similar coronal structures seen in the ultraviolet (UV) emerging from a coronal hole were found to be the same object as the white-light plumes (DeForest and Gurman, 1998). In the corona, the plume's general morphology has been related to the global structure of the polar magnetic field. From white-light observation, plumes show a super-radial expansion with altitude (DeForest *et al.*, 1997; Feng *et al.*, 2009). Close to the surface, plumes appear to be roughly linear structures, which converge to points on the solar-rotation axis more than half the way between the center of the Sun and the poles as expected for near dipole magnetic field lines. It has been suggested by Wang and Sheeley (1995) and Wang (1998) that plumes are formed by reconnection between small magnetic dipoles and larger unipolar flux concentration. However, they are observed above unipolar and bipolar flux concentrations (Wang, 1998).

[1] Max Planck Institute for Solar System Research, Max-Planck-Str. 2,
37191 Katlenburg-Lindau, Germany
email: depatoul@mps.mpg.de  email: inhester@mps.mpg.de
email: Wiegelmann@mps.mpg.de
[2] Institute for Geophysics and Extraterrestrial Physics, Technische
Universität Braunschweig, 38106 Braunschweig, Germany
[3] Purple Mountain Observatory Chinese Academy of Sciences,
210008 Nanjing, China
email: lfeng@pmo.ac.cn

The observations of EUV plumes actually show their projection on the plane of the sky. Therefore, it is impossible to determine their 3D geometry just from 2D observations. Indeed, plume observations are a line-of-sight integration of optically thin, straight objects. Moreover, plumes appear to be quite diffuse and not well confined to a clear footpoint, and therefore plumes and inter-plume regions are not sharply bounded. Finally, the lifetime and the 3D cross-section of a plume are poorly known. Llebaria *et al.* (1998) and DeForest, Lamy, and Llebaria (2001) report two different temporal ranges, the first authors found a lifetime of one–three days and the second found a lifetime of a week and even longer.

Different assumptions of the 3D plume cross-section shape and plume distribution have been proposed in the literature: DeForest *et al.* (1997) defined plumes as objects with a near-elliptical cross-section having a diameter of about 30 Mm growing with height. These objects have a lifetime of the order of a week and occasionally longer. Llebaria, Saez, and Lamy (2002) propose a fractal structure for the plume cross-section, which would explain the large variations of their thickness and spatial distribution. Wang and Sheeley (1995) suggest that plumes are closely related to the network activity and they claim a curtain- or sheet-like shape also proposed by Gabriel, Bely-Dubau, and Lemaire (2003). Gabriel *et al.* (2005) suggested the existence of two different plume populations having distinct geometrical forms: the classical *beam plumes* having a near-elliptical cross-section with a lifetime of about a week and the *curtain plumes* that would be only visible when the curtain is sufficiently aligned with the line-of-sight. This geometric effect limits the detectability of a curtain plumes to about one–three days. Since plumes are closely related to network activity, Gabriel *et al.* (2009) proposed that the supergranular network provides the required spatial distribution. Barbey *et al.* (2008) have used tomographic techniques for the reconstruction of North Pole plume observations showing a network pattern as proposed by Gabriel *et al.* (2009).

The measurements of temperature and density in plumes have been studied by Wilhelm and Bodmer (1998) and Wilhelm (2006) who found electron temperatures below one MK inside plumes. Considering the beam-plume geometry, they suggest that their densities are five times higher than in the nearby inter-plume corona. The outflow velocities in plumes have been determined using the Doppler-dimming technique with both SUMER and UVCS instruments (Teriaca *et al.*, 2003; Gabriel, Bely-Dubau, and Lemaire, 2003; Raouafi, Harvey, and Solanki, 2007). The conclusions drawn by these articles argue for different scenarios: Teriaca *et al.* (2003) found only low velocities and concluded that plumes cannot be considered as the source of the fast solar wind. This was also confirmed by Feng *et al.* (2009) who corrected the observed Doppler shift by the angle of the reconstructed 3D plume axis with the line-of-sight. Whereas Gabriel, Bely-Dubau, and Lemaire (2003) argue that plumes have a greater outflow speeds than in the inte-plume regions. Raouafi, Harvey, and Solanki (2007) agree with the low velocities in plume at low altitudes below $2\,R_\odot$ and show that it increases with height and reaches the inter-plume values above roughly $3-4\,R_\odot$.

A review on morphology, dynamics and plasma parameters of plumes and inter-plume regions is given by Wilhelm *et al.* (2011) and a general review taking all aspects of coronal plumes into account (Poletto, 2011).

In this article, EUV plumes are analyzed from the SOHO/EIT and STEREO/EUVI images. They are best seen in the Fe IX line at 171 Å. In Section 2, we present the dataset used to analyze EUV polar plumes. In Section 3, a method capable of automatically identifying and extracting some characteristics of solar polar plumes in 171 Å close to the limb at $1.01-1.39\,R_\odot$ is described. This method is basically a multiscale analysis using a combination of Hough and wavelet transform. It is applied to the full dataset presented in the first section. The result is a time series, which will be analyzed and discussed. Finally, Section 4 concentrates on stereoscopic and tomographic methods adapted for plume analysis. Cross sections and distributions of plumes and inter-plume regions over the Pole will be discussed.

## 2. Data

The plumes were observed by the two *Extreme Ultra Violet Imagers* (EUVI) belonging to the instruments SECCHI-A and SECCHI-B onboard the two STEREO spacecraft A and B, respectively (Howard *et al.*, 2008) and the *Extreme ultraviolet Imaging Telescope* (EIT) onboard the SOHO spacecraft (Delaboudinière *et al.*, 1995). The sequence of images considered here was taken at 171 Å and during the period from 1 November 2007 00:00 UT to 1 December 2007 00:00 UT. Although all three instruments are imagers in the extreme UV, many differences exist. The first data from EIT were taken in 1995, compared to EUVI-A and B where the first images were recorded at the beginning of 2006. The image quality of EIT has decreased since then. Additionally, there exist some differences in the technology of the instruments. For example, EUVI is a full-disk imager with a $1.7\,R_\odot$ field of view while EIT has a $1.5\,R_\odot$ field of view. The resolutions are respectively 1.6 arcsec pixel$^{-1}$ and 2.5 arcsec pixel$^{-1}$ for EUVI and EIT. The temporal cadences for the EUVI images at 171 Å were on average two minutes. Whereas the cadence for EIT images at 171 Å were six hours except for the end of November 2007, when there was a campaign with a special cadence of about seven minutes. Figure 1(top) is an example of a typical image from the EUVI-A instrument zoomed on the North Pole of the Sun.

Some pre-processing was applied to the images: Firstly, the standard calibration programs are used such as secchi_prep and eit_prep from the SolarSoft library (Freeland and Handy, 1998) for EUVI images and EIT images, respectively. Secondly, a pre-processing is applied in order to remove cosmic rays. It employs a basic median filter with a box of five pixels iterated three times over the image. Cosmic-ray pixels are considered to have values higher than $3\sigma$ of the median-filtered image neighborhood.

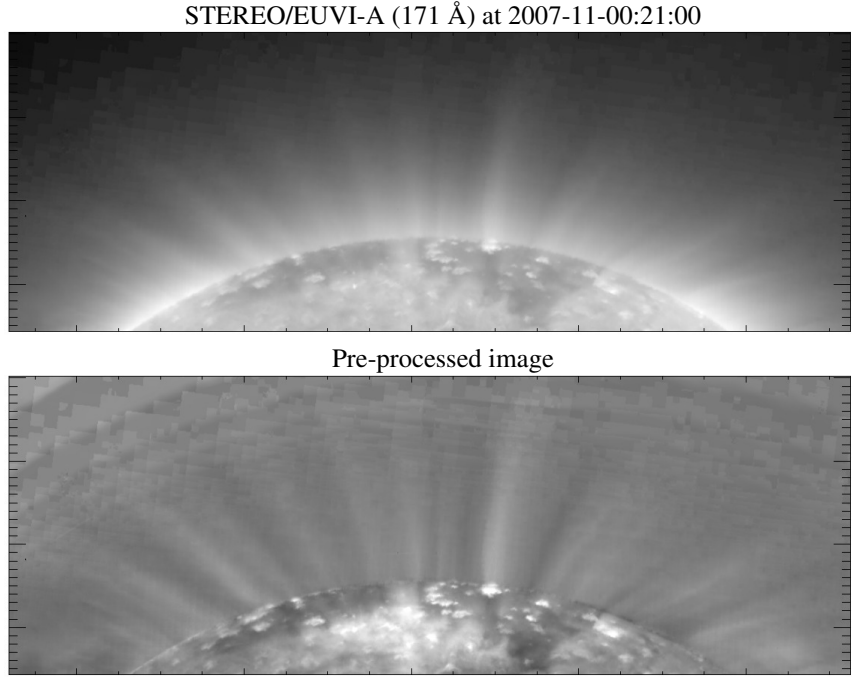

**Figure 1.** Image of the solar corona during the minimum of the solar cycle, EUV plumes appear as elongated straight faint columns. Top: Calibrated image represented logarithmically. Bottom: Same image enhanced following the procedure (1).

These pixels values are subsequently compared with those of the previous image to clearly identify the pixel as a cosmic-ray enhancement and not as a persistent local enhancement due to a solar event. Then pixels identified as cosmic-ray hits are simply replaced by the local median value. Finally, since plumes appear faintly in the image, applying a contrast-enhancement technique is necessary. For this purpose, a background image $[B(\boldsymbol{x})]$ is accumulated from a set of images $[I(\boldsymbol{x};t)]$ observed at time $t$ according to the following prescription:

$$B(\boldsymbol{x},t) = \min_{t' \in [t-15\,\text{day},\,t+15\,\text{day}]} \{\,\text{median}_{t'' \in [t'-12\,\text{hour},\,t'+12\,\text{hour}]}\, I(\boldsymbol{x},t'')\,\}.$$

From each image $[I(\boldsymbol{x};t)]$ a new enhanced image $[\tilde{I}(\boldsymbol{x};t)]$ is processed as

$$\tilde{I}(\boldsymbol{x};t) = \begin{cases} \frac{I(\boldsymbol{x};t)-B(\boldsymbol{x})}{B(\boldsymbol{x})} & \text{for } \boldsymbol{x} \text{ such as } I(\boldsymbol{x};t) \geqslant 0, \\ -1 & \text{for } \boldsymbol{x} \text{ such as } I(\boldsymbol{x};t) < 0. \end{cases} \quad (1)$$

A slightly more sophisticated background subtraction and contrast enhancement for different scale features using Haar wavelets has been proposed by Stenborg and Cobelli (2003) and Stenborg, Vourlidas, and Howard (2008). However, by using the Haar wavelet, the image intensity is locally modified and it cannot be ruled out that the intensity of plumes is changed differently or that the plume is even suppressed entirely.

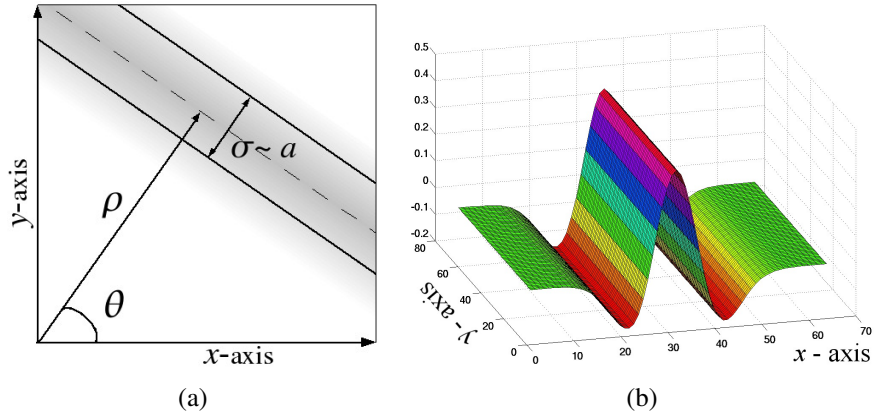

**Figure 2.** (a) Geometric illustration of the Hough-wavelet parameters $\theta \in ]-\pi/2, \pi/2]$, $\rho \in [-\rho_{\max}, \rho_{\max}] \subset \mathbb{R}$ and $a \in \mathbb{R}_0^+$. The parameter $a$ is proportional to the cross-section width $\sigma$. (b) An example of a wavelet (2) generated from the Mexican Hat mother wavelet (4) with the parameters $a = 8$, $\rho = 32$ and $\theta = 0$.

Keeping this background subtraction method simple, our procedure gives, nevertheless, a satisfactory enhancement of the plumes (Figure 1, bottom).

## 3. Plume Identification and Temporal Evolution in 2D

In this section, we will establish a method able to automatically identify plumes in a series of EUV images. In this method, a certain number of parameters will characterize each identified plume, allowing to follow their temporal evolution and to calculate their 3D locations and orientations on the surface of the Sun.

3.1. Hough-Wavelet Transform

In a first approximation, a projected polar plume appears close to the limb as a straight column with a certain width as shown in the Figure 1. A suitable method for extracting straight, elongated objects is the Hough transform (Toft, 1996). Basically, the Hough transform converts a line in an image into a point characterized by two parameters in Hough space: the inclination [$\theta$] and the distance [$\rho$] from the origin Figure 2(a). In this way, the difficulty of the global detection problem in an image is reduced to a more easily solvable local-peak detection problem in Hough space. In solar physics, the Hough transform has been used for the first time by Llebaria and Lamy (1999) to detect plumes in white-light coronograph images, then later by Robbrecht and Berghmans (2004) for the automated recognition of coronal mass ejections (CMEs) in near-real-time data.

Plume features have a certain width and the basic Hough transform is not well adapted to faint, finite-width objects. Therefore, we propose here to combine the Hough transform with the wavelet transform, which we call the *Hough-wavelet transform*. A similar extension of the Hough transform, but with a different normalization, was also introduced in image processing as the *ridgelet transform* (Candès and Donoho, 1999).

In continuous Hough-wavelet analysis, a 2D signal $[f]$ is analyzed with a smooth-function $[\psi]$ with sufficient decay and zero mean from which a family of wavelets $[\psi_{\rho,a,\theta}]$ is generated by elementary operations, *i.e.* rigid translation by a distance $\rho$, rotation by an angle $\theta$, and dilatation by a factor $a$. These three elementary transformations yield the wavelets,

$$\psi_{\rho,a,\theta}(\boldsymbol{x}) = \frac{1}{a} \, \psi\left(\frac{\boldsymbol{u}_\theta \cdot \boldsymbol{x} - \rho}{a}\right) \quad \text{with} \quad \boldsymbol{u}_\theta = (\cos\theta \quad \sin\theta) \,, \qquad (2)$$

where $a \in \mathbb{R}_0^+$ is the scale and $a^{-1}$ is a normalization factor. $\rho \in [-\rho_{\max}, \rho_{\max}] \in \mathbb{R}$ is the distance between the center of $\psi_{\rho,a,\theta}$ and the origin. By convention, $\rho$ is negative for $\theta \in$ quadrant $II$ and $III$ and $\theta \in [-\pi/2, \pi/2[$ is the angle formed by the vector $\boldsymbol{u}_\theta$ and the $x$-axis. Figure 2(a) describes the Hough-wavelet parameters in image-space. The 2D signal $f$ is then analyzed with these wavelets by transforming it to a function of the three variables $(\rho, a, \theta)$:

$$H(\rho, a, \theta) = \frac{1}{a} \int_{\mathbb{R}^2} f(\boldsymbol{x}) \, \psi\left(\frac{\boldsymbol{u}_\theta \cdot \boldsymbol{x} - \rho}{a}\right) \mathrm{d}\boldsymbol{x} \,, \qquad (3)$$

where we assume $\psi$ to be real. The choice of the normalization factor $a^{-1}$ in (2) has the advantage of giving more weight to the small scales in (3). This produces an enhancement at the high-frequency part of the signal, and thus emphasizes its singularities. For mathematical reasons, $a^{-1/2}$ is often selected as normalization in order to make the wavelet transform unitary (Antoine *et al.*, 2004). Note that if we choose a "degenerate" mother wavelets, *i.e.* the Dirac function $[\delta(\boldsymbol{x})]$, we obtain the basic Hough transform again (Holschneider, 1993).

In this work, we propose the Mexican Hat wavelet as the mother wavelet, which is the Laplacian of a Gaussian:

$$\psi^{\mathrm{MH}}_{(\boldsymbol{x})} = -\Delta \exp\left(-\frac{|\boldsymbol{x}|^2}{2}\right) = (2 - |\boldsymbol{x}|^2) \, e^{-\frac{|\boldsymbol{x}|^2}{2}}.$$

This wavelet is quite robust to noise (since it has two vanishing moments) and it is easy to compute numerically (the function and its Fourier transform are real). The family of wavelets generated from the Mexican Hat mother wavelet are according to Equation (2),

$$\psi^{\mathrm{MH}}_{\rho,a,\theta}(\boldsymbol{x}) = -\frac{1}{a}\Delta \exp\left(-\frac{|\boldsymbol{u}_\theta \cdot \boldsymbol{x} - \rho|^2}{2a^2}\right) \,. \qquad (4)$$

Figure 2(b) shows an example of this wavelet translated by $\rho = 35$, dilated by a factor $a = 8$, and not rotated $[\theta = 0]$. This method yields a multi-scale analysis with a good signal to noise ratio. Moreover, the Hough-wavelet coefficients (3) contain all of the information of the original image (Torrésani, 1995; Antoine *et al.*, 2004). Therefore this transform has the property of being invertible and allows a perfect backward projection into the image space.

### 3.2. Hough-Wavelet Transform on Synthetic Data

The Hough-wavelet transform can be analytically calculated for a Gaussian line signal with a certain width [$\sigma$] and an amplitude [$A$],

$$f(\boldsymbol{x}) = A \exp\left(\frac{-y^2}{2\sigma^2}\right) . \tag{5}$$

The Hough-wavelet coefficients of this signal maximizes when the parameters $\rho$, $a$, and $\theta$ match in position, orientation, and scale with the Gaussian signal $f$ in the image. In our example, these parameters are $\rho = \rho_\mathrm{M} = 0$ and $\theta = \theta_\mathrm{M} = \pi/2$. The scale $a = a_\mathrm{M}$ is proportional to the width [$\sigma$] of the Gaussian and the maximum coefficient value is proportional to the amplitude of the Gaussian [$A$] and its width [$\sigma$]:

$$\begin{aligned} H(\rho, a, \theta_\mathrm{M}) &= \int_{\mathbb{R}^2} f(\boldsymbol{x})\, \psi^{\mathrm{MH}}_{\rho, a, \theta_\mathrm{M}}(\boldsymbol{x})\, \mathrm{d}\boldsymbol{x} \\ &= -\frac{A a \sigma \sqrt{2\pi}}{(a^2 + \sigma^2)^{\frac{3}{2}}} \left(\frac{\rho^2}{(a^2 + \sigma^2)} - 1\right) \exp\left(\frac{-\rho^2}{2(a^2 + \sigma^2)}\right) , \end{aligned} \tag{6}$$

with $\rho = \rho_\mathrm{M} = 0$ yield

$$H(\rho_\mathrm{M}, a, \theta_\mathrm{M}) = \frac{A a \sigma \sqrt{2\pi}}{(a^2 + \sigma^2)^{\frac{3}{2}}} , \tag{7}$$

which has a maximum at scale $a = a_\mathrm{M}$:

$$a_\mathrm{M} = \sqrt{2}\sigma . \tag{8}$$

At this scale [$a = a_\mathrm{M}$] the Hough-wavelet coefficients have a value of

$$H(\rho_\mathrm{M}, a_\mathrm{M}, \theta_\mathrm{M}) = \frac{2\sqrt{\pi}}{3\sqrt{3}} \frac{A}{\sigma} . \tag{9}$$

The Hough-wavelet coefficients are therefore proportional to the amplitude [$A$] of the Gaussian and inversely proportional to the Gaussian width [$\sigma$]. Figure 3(a) shows a synthetic image of three plumes with Gaussian cross-sections of width $\sigma = 2$ for the left and right plumes and $\sigma = 1$ for the center plume. All of these plumes have the same amplitude $A$. The Hough-wavelet coefficients are shown in Figure 3(b). They are computed for the scales $a = 1.1, .., 10$ taken with logarithmically equidistant steps. For each position [$(\rho, \theta)$], the characteristic scales [$a_\mathrm{M}(\rho, \theta)$] are then computed and

$$H_\mathrm{M}(\rho, \theta) \equiv H(\rho, a_\mathrm{M}(\rho, \theta), \theta) = \max_a \left[ H(\rho, a, \theta) \right] . \tag{10}$$

Figure 3(a) displays a cut at $\theta = 0$ showing that the middle plume has a coefficient twice larger than the two other plumes, which is expected from Equation (9).

J. de Patoul *et al.*

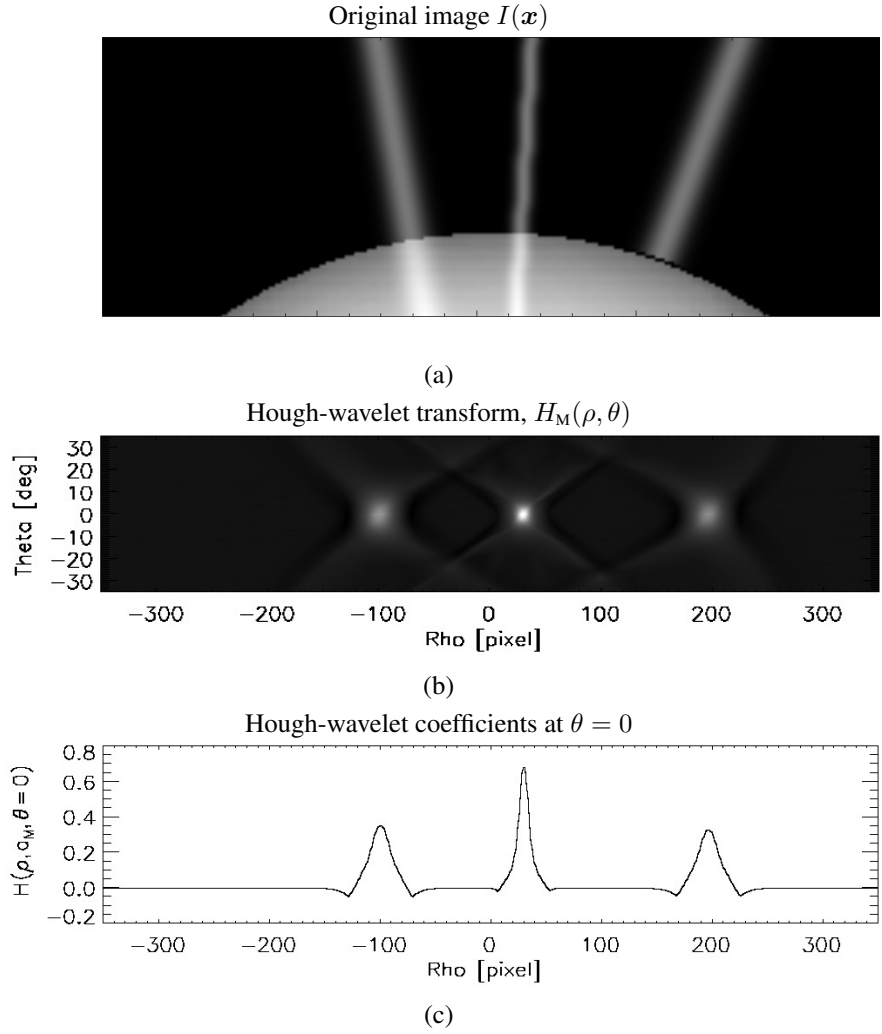

**Figure 3.** In (a) an image with three synthetic plumes with Gaussian cross-sections. Its Hough-wavelet transform is shown in (b) at the selected scale $a_{\rm M}$ computed separately for each pixel. (c) is a cut of panel (b) at $\theta = 0$ showing that the middle plume has a coefficient two times larger than the two other plumes.

### 3.3. Hough-Wavelet Transform on Real Data

Before we can apply our transform to real data, we map the image of the solar polar region to cylindrical coordinates in order to get rid of edge effects in the image: $\check{I}(r, \phi; t) = I(x(r, \phi), y(r, \phi); t)$. Here $r$ is the distance from the projected solar center and $\phi$ is the angle with respect to the projected solar-rotation axis *i.e.*:

$$\begin{cases} r = \sqrt{x^2 + y^2}\,, \\ \phi = \arctan\left(\frac{x}{y}\right)\,. \end{cases} \tag{11}$$

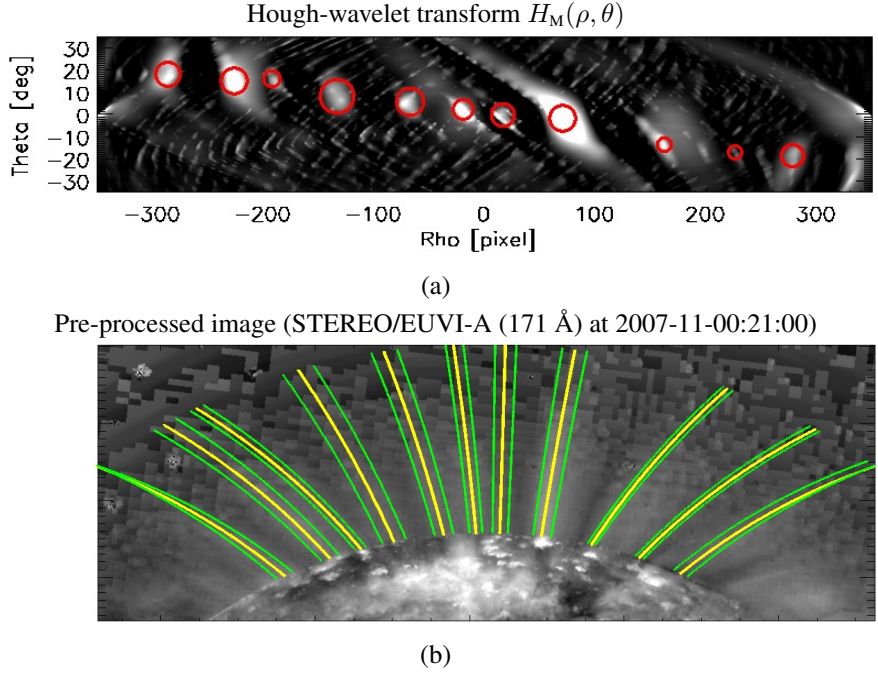

**Figure 4.** (a) Red circles correspond to intensity maxima over different scales of the Hough-wavelet coefficients. (b) The backward transformation of polar plumes on the coronal image. The yellow lines correspond to local maxima in the Hough space and in green the computed width of the plume.

See Figure 5. The obtained image $\breve{I}(r,\phi;t)$ is restricted to $\phi \in [-35°, 35°]$ and $r$ from $1.01\,R_\odot$ to $1.39\,R_\odot$. The signal $\breve{I}(r,\phi;t)$ is then analyzed by the Mexican Hat wavelet (4) with the Hough-wavelet transform (3) for scales $a \in [2.00, 10.00]$ varied in logarithmically equidistant steps:

$$H(\rho, a, \theta) = \int_{-35°}^{35°} \int_{1.01\,R_\odot}^{1.39\,R_\odot} \breve{I}_t(r,\phi)\, \psi_{\rho,a,\theta}^{\mathrm{MH}}(r,\phi)\, dr\, d\phi\,. \quad (12)$$

A point $(\rho, \theta)$ in Hough-space corresponds to a line in $(r, \phi)$-space defined by

$$\begin{pmatrix} \cos\theta \\ \sin\theta \end{pmatrix} \cdot \begin{pmatrix} \phi\,\frac{\rho_{\max}}{\phi_{\max}} \\ r - r_c \end{pmatrix} = \rho\,, \quad (13)$$

where $r_c$ is the center of the Hough-wavelet coordinate frame (Figure 5), here $r_c = 1.19\,R_\odot$ and $\rho_{\max} = 350$ pixel, $\phi_{\max} = 35°$. The point $(\rho, \theta)$ corresponds to a curve in image space, which is implicitly defined by Equation (13) with Equation (11) inserted for $r$ and $\phi$. Then for each position $[(\rho, \theta)]$, the maximum of $H(\rho, a, \theta)$ over the scales $a$ is then determined by Equation (10).

An example of the Hough-wavelet transform is shown in Figure 4(a). The centers of the red circles correspond to the local maxima, and the radii are proportional to the optimum scale $a_{\mathrm{M}}$. The equivalent Gaussian width of the selected plumes is $a_{\mathrm{M}}/\sqrt{2}$.

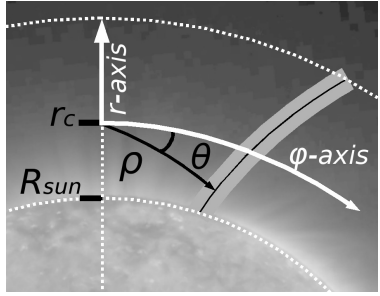

**Figure 5.** Hough-wavelet coordinates for the solar polar region in cylindrical coordinates.

The maxima are aligned along an inclined line

$$\theta \cong -0.15\, \rho\, \frac{\deg}{\text{pix}}, \qquad (14)$$

due to the super-radial inclination of the plumes.

As the transformation is invertible, the backward projection is unique. One point with coordinates $(\rho, a, \theta)$ in the Hough-wavelet space corresponds to one column with a width proportional to $a$. In Figure 4(b), the yellow lines indicate the ridge of the detected plume and the green curves represent the width of the plume given by the scale $a_M/\sqrt{2}$. The yellow curves are not straight because the Hough-wavelet transform was applied to the image after the mapping to cylindrical coordinates. The respective curve is defined implicitly by Equation (13) with Equation (11) inserted for $r, \phi$.

### 3.4. Time Series and Sinogram

In this paragraph, the Hough-wavelet procedure described above will be applied to the full series of data presented in Section 2. The Hough-wavelet coefficients from the whole time series of images can be concisely represented by a *sinogram*. A similar time–intensity diagram to study plumes was proposed for the first time by Llebaria *et al.* (1998) using the basic Hough coefficients as input, while here we use the Hough-wavelet coefficients. Formally, the sinogram is computed as follows

$$S(\rho;\, t) = \int_{-35°}^{35°} H_M(\rho, \theta; t)\ \ W(\rho, \theta)\ \ d\theta\,. \qquad (15)$$

Here $W(\rho, \theta)$ is a Gaussian weight function, which gives more weight to the area in Hough-wavelet space corresponding to the inclination line (14), along which the plume signal are lined up, see Figure 4(a),:

$$W(\rho, \theta) = \exp\left(\frac{-\left(\theta + \rho\ 0.15\ \frac{\deg}{\text{pix}}\right)^2}{2\sigma^2}\right).$$

Figure 6 shows the resulting sinogram (15) obtained from real data: EUVI-B in (a), EIT in (b) and EUVI-A in (c). The black rows have been filled for the missing data in

STEREO/EUVI-B

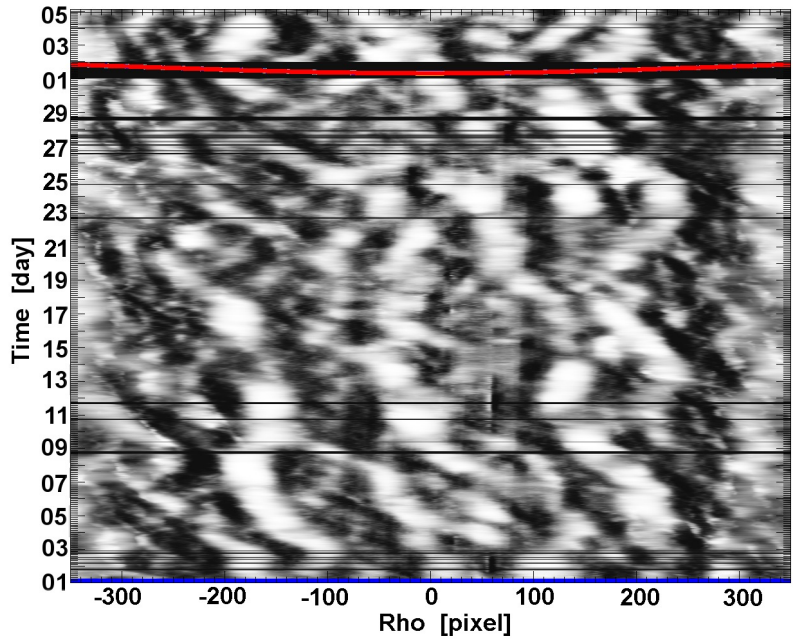

(a)

SOHO/EIT

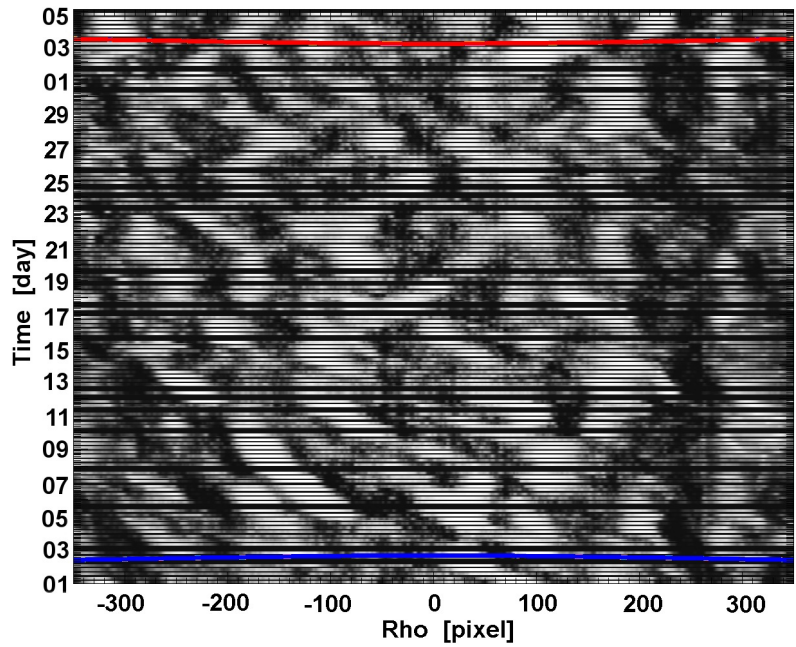

(b)

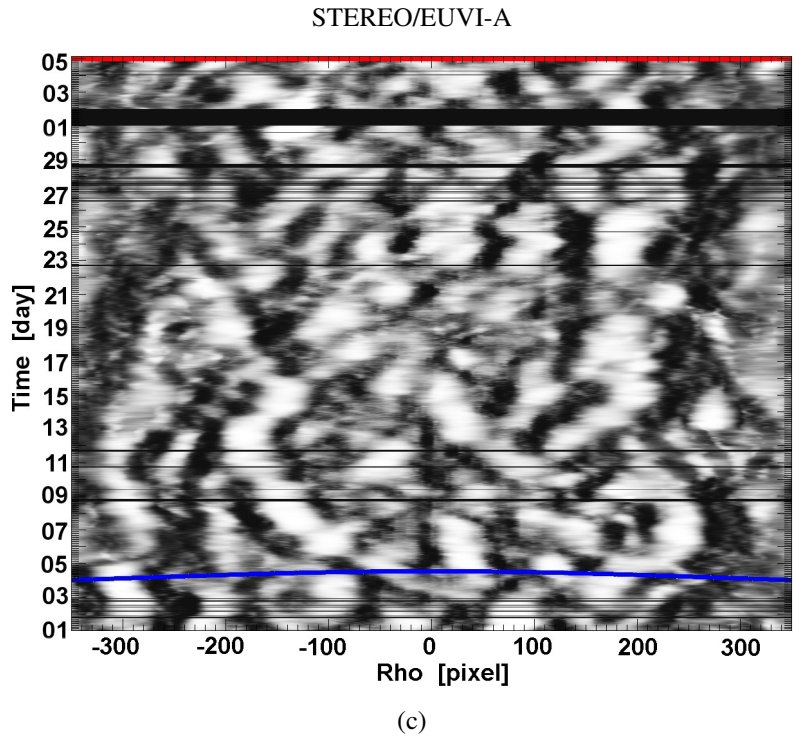

(c)

**Figure 6.** Sinogram from 1 November to 5 December 2007. This period of time corresponds to 360° at the Pole. The black lines correspond to missing data. The blue and red curves correspond to the view direction of the STEREO spacecraft B and A respectively on 1 November 2007.

order to present the observations equidistantly in time. Note that EIT has a shorter image cadence. The view direction on 1 November 2007 of the STEREO spacecraft B and A are drawn in blue and red respectively. On these diagrams, plumes should trace out sections of sinusoids according to their life time and position on the solar surface. The red and blue curves indicate where the observation in one instrument is mapped into the singogram of the other instruments. Due to the different vie angles the curves are essentially time shifted according to the solar rotation.

## 4. Reconstruction of the 3D Structure

The sinogram contains all the plume data observed. It is the starting point for different methods of analysis: the sinogram from a single instrument over the entire time range can be used for tomography, or it can be used from two instruments at a given time as input for stereoscopy.

4.1. Tomography

The basic problem of classical tomography is the reconstruction of a 2D image from a set of 1D projections on different viewing directions. In this article, the 1D projection is

Sinogram from 3D model

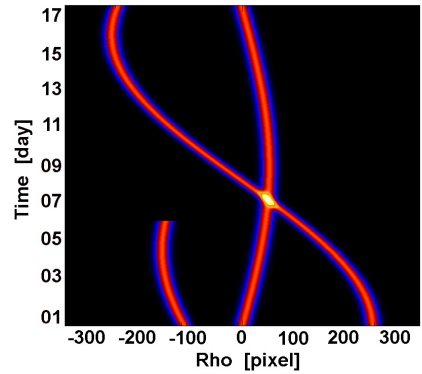

(a)

Classical tomographic reconstruction

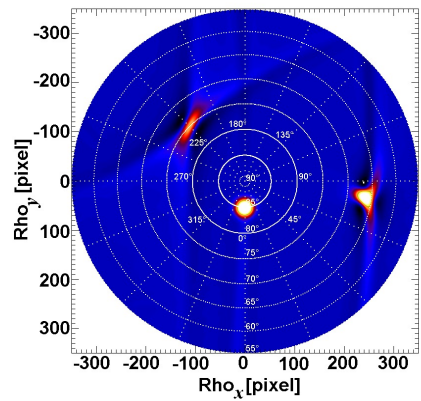

(b)

Tomographic reconstruction including differential rotation the Hough-wavelet coefficients from the sinogram (15) at a fixed time. The tomography method used here is the *filtered back-projection reconstruction* described by Natterer (2001). In this approach, each of the 1D views is convolved with a filter before the back-projection operation. We use the filter function suggested by Natterer (2001). For $\rho$ discretized at equidistant steps $\rho_n = -N/2, ..., N/2$ with $N = 700$ and $\epsilon \in [0, 1]$,

$$F(\rho_n) = \frac{N^2}{16\pi} \begin{cases} \frac{1}{4} - \frac{\epsilon}{6}, & \text{if } \rho_n = 0, \\ \frac{-\epsilon}{\pi^2 \rho_n^2}, & \text{if } \rho_n \neq 0 \text{ and } \rho_n \text{ is even}, \\ \frac{-(1-\epsilon)}{\pi^2 \rho_n^2}, & \text{if } \rho_n \neq 0 \text{ and } \rho_n \text{ is odd}. \end{cases} \qquad (16)$$

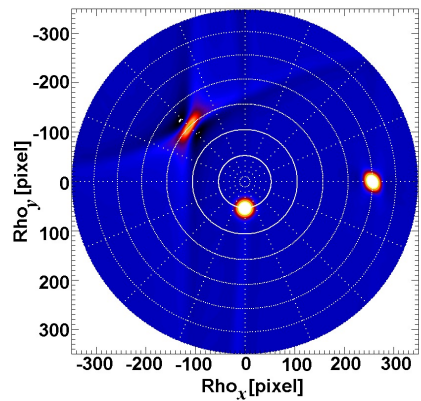

(c)

**Figure 7.** (a) Sinogram computed for the synthetic 3D model to which Figure 3(a) shows a projection. One of the events lasts only for the first 60°. (b) Classical filtered back-projection computed for 180° of the 3D model. The model turned clockwise and the view direction was downward. The result is shown from 90° latitude in the center to 55° latitude. (c) Filtered back-projection computed for the same model as (b) with differential rotation applied as in Equation (18).

The Hough-wavelet coefficients for a given time are first convolved with the filter (16), then subsequently the back-projection is applied.

In order to properly interpret the results of the procedure applied to real data, we briefly discuss the outcome from a simulated model of three Gaussian-shaped plumes rooted at latitude $85°$, $65°$, $75°$ and longitude $0°$, $90°$, $225°$ respectively. A projection of this model is shown in Figure 3(a). From this model, we generated the sinogram (Figure 7(a)) from the Hough-wavelet transform (15) taking into account two discrepancies of the model from the assumption of classical tomography. Classical tomography is based on two implicit assumptions: solid rotation and stationarity of the objects while they rotate. Both assumptions do not hold well in the case of plume observations. The solar corona does not rotate as a rigid body, but shows an angular velocity varying with latitude. Here we assume a differential-rotation law from Cox (2000),

$$\omega(\lambda) = 14.713 - 2.396 \sin^2(\lambda) - 1.787 \sin^4(\lambda) . \qquad (17)$$

Depending on the latitude, the time $t$ is between 17.09 days at the Pole and 14.63 days at $55°$. To see the effects of non-stationarity, the lifetime of the plume at latitude $65°$ was limited to a sixth of a full rotation.

Figure 7(b) shows the filtered back-projection of this synthetic model for the filter parameter $\epsilon = 1$ but without taking account of the differential rotation. The result corresponds to a 2D distribution of a density on a plane normal to the solar-rotational axis. The center of the image Figure 7(b) corresponds to the North Pole. The locations of the three model plumes are well retrieved from this ideal data set. The positions on the image, approximately reproduce the latitude and longitude of the plume foot-points (see relation (13) for the co-latitude), the sizes are proportional to the plume widths (8) and the intensities are also in agreement with Equation (9).

While the plume close to the center is positioned at the right location and corresponds to the shape used in the model to generate the data, the two other plumes do not show the circular cross-section given as input in the synthetic model. The plume at lower latitude also is miss-positioned of few degrees longitude due to the wrong rotation rate assumed for its latitude. If differential rotation is not taken into account, the reconstructed plume cross-section shows a typical deformation as for the low-latitude plume in Figure 7(b).

Our backprojection algorithm incorporates differential rotation (17) and the cross-section density $D(x, y)$ in the plane normal to the solar-rotation axis is calculated

according to

$$\mathrm{D}(x,y) = \sum_t \int_{-\infty}^{\infty} \delta\left(x\cos[\omega(\lambda)\,t] + y\sin[\omega(\lambda)\,t] - \rho\right)$$
$$\int_{-\infty}^{\infty} \mathrm{F}(\rho - \rho')\mathrm{S}(\rho',t)\,\mathrm{d}\rho'\,\mathrm{d}\rho\,, \qquad (18)$$

where $t$ is time, $\omega(\lambda)$ is given by Equation (17) and the latitude $\lambda = \pi/2 - \sqrt{(x^2+y^2)}$. Figure 7(c) shows the back-projection with differential rotation applied. The effect of a limited lifetime on the tomographic reconstruction can be studied with the second plume of our model at latitude $65°$. Again, its reconstructed cross-section appears at about the right position but heavily distorted. The major axis of the reconstructed shape is aligned along the mean direction towards the observers during the lifetime of the plume. Hence, reconstructions from the data of different spacecraft should yield elongated cross-sections but with different orientations of the major axis.

Figure 8(a), (b), and (c) show the back-projection results calculated independently from EUVI-B, EIT, and EUVI-A data respectively from the left to the right. The density $\mathrm{D}(x,y)$ has been calculated from the North Pole to latitude $55°$N and for the images taken between 1 and 19 November 2007. The North Pole is located at the center of the image. All images are rotated to the Heliocentric Earth EQuatorial (HEEQ) coordinate system with Earth and SOHO located toward the bottom of the image, EUVI-A and B located at an angle of about $20°$ to the right and to the left from downwards, respectively. The separation between plumes with high intensity (yellow to white color on the image) and dark inter-plume regions (blue to black color) are clearly defined. However, the three images do not give an identical result. One reason is different sensitivity and signal-to-noise ratio of each of the three instruments. The other more important reason comes from the fact that the plume intensity varies considerably in time so that the plume configuration is changed already after a solar rotation of about $20°$. In November 2007, this time difference is about two days based on the rotation of the Pole and 1.5 days at $55°$ latitude. Hence, the data obtained by each spacecraft differ considerably.

4.2. Stereoscopy

In addition to its sensitivity for faint plume signals, another advantage of the Hough transform is that the Hough-space coordinates $(\rho, \theta)_I$ for $I = A, B$ or $E$ of plumes detected in the images of EUVI-A, EUVI-B, and EIT respectively, yield directly the plume 3D axis. This axis is given by the intersection of two planes $\pi_A \cap \pi_B$, $\pi_E \cap \pi_B$ or $\pi_A \cap \pi_E$ each of which is readily constructed: see Figure 9. In our coordinate system, the planes $\pi_I$ are given by

$$\pi_I(s,t) = P_I + s\,(\boldsymbol{e}_I + t\,(\boldsymbol{h}_I x_1 + \boldsymbol{v}_I y_1)\Lambda + (1-t)(\boldsymbol{h}_I x_2 + \boldsymbol{v}_I y_2)\Lambda)\,, \qquad (19)$$

where $\Lambda$ is the image conversion factor from pixels to arcsecs, $s$ and $P_I$ are in HEEQ coordinates with $P_I$ the satellite position EUVI-A, B or EIT, $\boldsymbol{v}_I$ and $\boldsymbol{h}_I$ are the 3D unit vector directions along the edges of the image and $\boldsymbol{e}_I$ is the image view direction

STEREO/EUVI-B

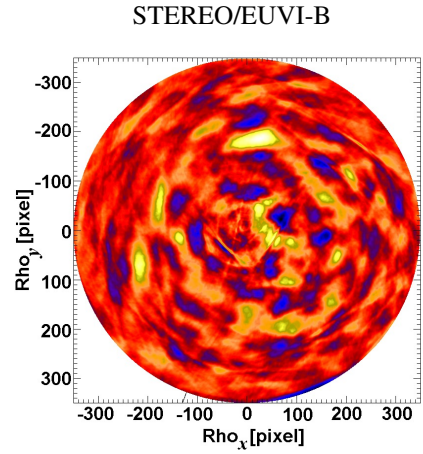

(a)

SOHO/EIT

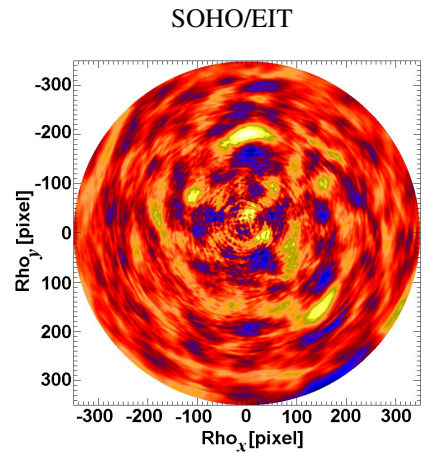

(b)

STEREO/EUVI-A

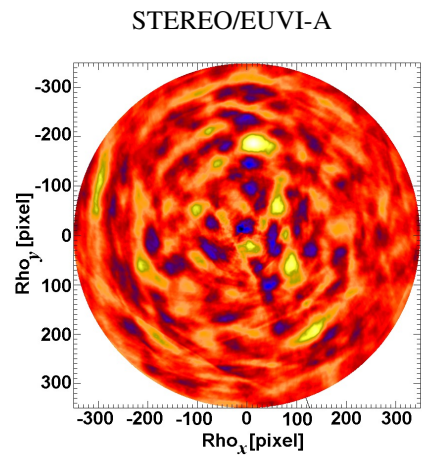

(c)

**Figure 8.** Filtered back-projections of the EUVI-B in (a), EIT in (b), and EUVI-A (c) sinograms, from North Pole up to lat. 55°N. North Pole is at the center of each image. In the direction straight downwards is the SOHO spacecraft. In the direction downwards left is the EUVI-B spacecraft at about $-20°$ longitude, and EUVI-A is downwards right at about $+20°$ longitude.

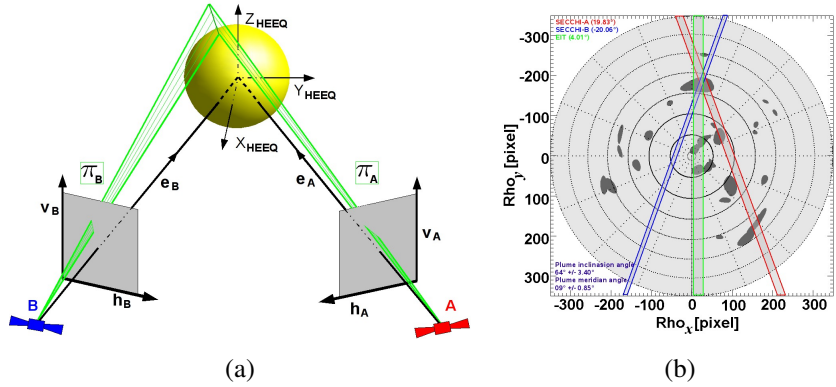

**Figure 9.** (a) Sketch of the Sun and instruments EUVI-A, EUVI-B in HEEQ coordinate. A plume located on the Sun is at the intersection of the two planes $\pi_A$ and $\pi_B$ and the solar surface. (b) View of the North Pole of the Sun in the same coordinates as Figure 8. The intersection of the strips gives the polygon within which the plume is concentrated.

perpendicular to $\boldsymbol{v}_I$ and $\boldsymbol{h}_I$. $(x_1, y_1)$ and $(x_2, y_2)$ are two arbitrary but different points to be calculated from Hough-space coordinates of the plume, $(\rho_I, \theta_I)$ in Equation (12):

$$\arctan\left(\frac{y_i}{x_i}\right) \cos\theta_I + \left(r_c - \sqrt{x_i^2 + y_i^2}\right) \sin\theta_I = \rho_I,$$

for $i = 1, 2$. Each intersection $\pi_A \cap \pi_B$, $\pi_E \cap \pi_B$ and $\pi_A \cap \pi_E$ yields a 3D plume axis, which ideally are all identical for a given plume (Figure 9(a)). The intersection of the axis with the solar surface then yields the footpoint position of the plume.

Additionally, the Hough-wavelet procedure gives the scale parameter $a_M$ (8), which is a measure of the apparent width of the plume. Given these widths on either side of the planes $\pi_I$, the intersection of a pair of planes gives an intersection region with the shape of a parallelepiped. For the three pairs from the three instruments A, B, and E, the intersection area shrinks to a polygon or is no-existent if the correspondence of the plumes in the three images was not selected correctly.

An example is shown in Figure 9(b) for a plume detected in EUVI-A, B, and EIT images. The colored lines (red, blue and green) are the view directions of EUVI-A, B, and EIT respectively. The distance between these pairs of lines corresponds to the widths of the boxes and therefore to the respective apparent width of the plume. The intersection of the strips bounded by the parallel colored lines indicates the area where the plume has to reside. Adding the third view point, EIT reduces this area compared with what we obtain from STEREO alone. The correspondence problem of an object in images taken from different view points is discussed by Inhester (2006). The necessary intersection from three views gives us an effective consistency check of the selected correspondence. Here, we have in addition cross checked the resulting plume positions with the tomography results. In Figure 9(b), the dark-gray patches represent the local maxima extracted from images Figure 8(a), (b), and (c).

## 5. Discussion

Our tomographic results show that plumes can have all types of cross-section shapes from elliptical with small eccentricity to really extended shapes. In order to investigate whether the shape is due to plume itself or is an artifact due to its short lifetime, we can check the lifetime of the plume by looking at its mark in the sinogram. Here we will discuss a few examples. There are enumerated in Figure 10:

*i*) The first case is the plume located at $(\rho_x, \rho_y) = (110, 200)$ corresponding to $(lon, lat) = (26°, 66°)$ (number 1 in Figure 10). This plume is visible in all three sinograms (Figure 10(b)) during the same period: starting 3 November and ending on 10 November. In this case, the plume cross-section obtained from tomography result is elongated due to its short lifetime compare to the time for half a rotation. Depending on the view direction, the elongation has a different orientation. The expected theoretical orientation is marked with a dashed line on the Figure 10(a).

*ii*) The second case is the plume located at $(\rho_x, \rho_y) = (0, -175)$ corresponding to $(lon, lat) = (180°, 73°)$ (number 2 in Figure 10). This plume is visible first in the EUVI-B sinogram on 2 November. It appears about two days later in the sinogram from EIT and about four days later in EUVI-A. In each sinogram it lasts for about eight days. This two-day and four-day delay corresponds exactly to the time for the corona at about 73° latitude to rotate from the EUVI-B view direction to SOHO ($\approx 20°$) and then to EUVI-A view direction ($\approx 40°$). In this case, we expect the true plume cross-section will be strongly elongated so that the plume becomes visible only when the major axis is aligned with the line-of-sight. The tomography result in Figure 10(a) shows an elongated cross-section with the same orientation for all three instruments.

*iii*) The third case is the plume located at $(\rho_x, \rho_y) = (95, 110)$ corresponding to $(lon, lat) = (47°, 74°)$ (number 3 in Figure 10). On the sinogram, we can see that the intensity of this plume varies with time. It can be interpreted as a series of plumes erupting at the same location. These variations may be due to recursive coronal jets located at the footpoint of the plume. These jets may be the responsible for the extended life cycle of these plumes as proposed by Raouafi *et al.* (2008). A further look is nevertheless necessary.

*iv*) Another case is the plume located at $(\rho_x, \rho_y) = (190, 90)$ which corresponds to $(lon, lat) = (292°, 84°)$ (number 4 in Figure 10). In this case the plume starts at the same time all the three sinograms, about 10 November. It is similar case in number 1 It lasts for about eight days and for some reason it is not well visible in the EIT instrument. In the tomographic reconstruction, the plume cross-section appears elongated, but the orientation of the elongation is typically rotated for EUVI-A and B due to the short lifetime.

*v*) The last case is the plume located at $(\rho_x, \rho_y) = (0, 30)$ corresponding to $(lon, lat) = (0°, 87°)$ (number 5 in Figure 10). In the sinogram of EUVI-B and EIT, we can see that it consists of successive bursts. During the first burst on 4 November at 02:25 UT, a jet was observed, which has been analyzed by Kamio *et al.* (2010). However, the plume is visible already before the occurrence of the burst. No X-ray jet was observed before the plume appeared. In this case, a jet as precursors of plumes, as suggested by Raouafi *et al.* (2008) did not occur. The lifetime of this plume was only about five days.

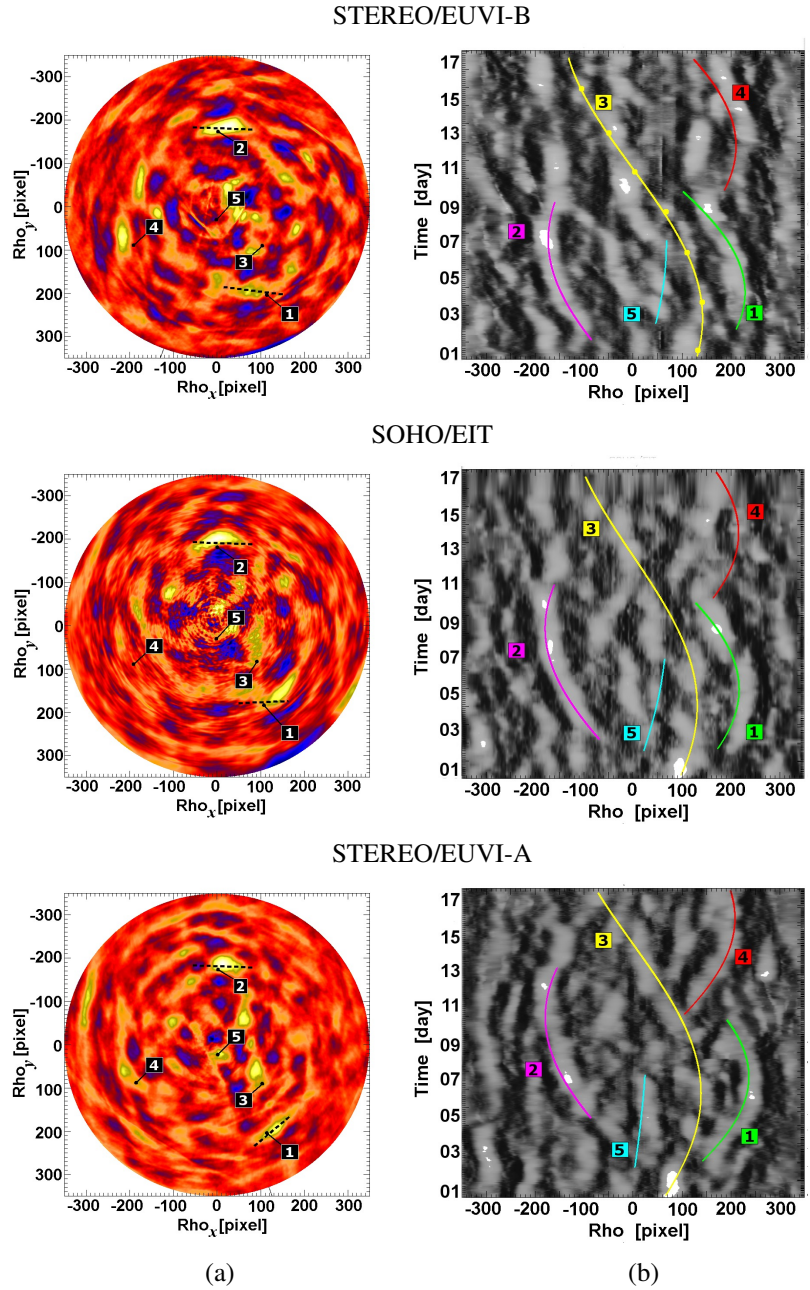

Figure 10. (a) Same results presented in Figure 8 from 1 November to 18 December 2007. Some selected plumes are marked by a number and are discussed in Section 5. (b) The corresponding sinograms used for the filtered back-projection results. The corresponding selected plumes in (a) are shown as colored lines with their associate number.

## 6. Summery and Conclusion

In this article we have shown that the Hough-wavelet transform is an appropriate method to identify the 2D projected plumes in EUV images and to study their temporal evolution. This method is also well suited as input for stereoscopic and tomographic 3D reconstruction. In particular, the Hough-wavelet parameters can be used directly for a stereoscopic calculation of the 3D location of a plume. For our tomographic reconstruction, we have taken the differential rotation of the solar surface into account to improve the reconstruction results. However, the results of the tomographic calculation still suffer severely from the non-stationarity of the plume structures. Roughly speaking, the tomographic reconstruction at a given longitude and latitude corresponds to the integral along the respective sinogram trace. Hence the inspection of the variation of the sinogram intensity along the respective trace helps to interpret the tomography results. Moreover, the fact that we can perform this analysis from the data of each spacecraft independently, helps greatly to cross check the interpretation. On the other hand, correspondence problem for stereoscopy is hardly solvable for plume observations given the number of detected plumes in the EUV images at a given time from the three view directions. Combining both methods, tomography and stereoscopy, can further help to identify the plumes, estimate their temporal variation and cross-section shape. So far we have studied few cases. Plumes number 1 and number 4 very probably were typical *beam plumes* with a localized cross-section. Plume number 2 we suspect to have a *curtain* shape with an elongated cross-section. This finding confirms the coexistence of *beam plumes* and *curtain plumes* already proposed by Gabriel *et al.* (2005). However, the tomography result shows that a broad range of intermediate configurations also exist. Plumes number 3 and number 5 were a more complex and seem to appear and disappear periodically at the same location with a lifetime of two – three days. The results show that the cross-section shape of the plume can be misinterpreted due to a limited plume lifetime.

**Acknowledgements**   The authors thank the SOHO/MDI and the STEREO/EUVI consortia for their data. SOHO and STEREO are joint projects of ESA and NASA. J. de Patoul was supported by the International Max Planck Research School on Physical Processes in the Solar System and Beyond run jointly by the Max Planck Society and the Universities Göttingen and Braunschweig. L. Feng was supported by the NSFC funding JY106111005. T. Wiegelmann was supported by DLR grant 50OC0501.